%% file: main.tex
\documentclass[conference]{IEEEtran}
\usepackage[inkscapelatex=false]{svg}

\IEEEoverridecommandlockouts
\usepackage{cite}
\usepackage{amsmath,amssymb,amsfonts}
\usepackage{algorithmic}
\usepackage{graphicx}
\usepackage{textcomp}
\usepackage{xcolor}
\usepackage{listings}
\usepackage{svg}
\usepackage{hyperref}

\def\BibTeX{{\rm B\kern-.05em{\sc i\kern-.025em b}\kern-.08em
    T\kern-.1667em\lower.7ex\hbox{E}\kern-.125emX}}

\begin{document}

\title{Towards a DSL to Formalize Multimodal Requirements\\

\thanks{This project is supported by the Luxembourg National Research Fund (FNR) PEARL program, grant agreement 16544475.}
}

\author{\IEEEauthorblockN{Marcos Gomez-Vazquez}
\IEEEauthorblockA{
\textit{Luxembourg Institute of Science and Technology}\\
Esch-sur-Alzette, Luxembourg \\
0000-0001-7176-0793}
\and
\IEEEauthorblockN{Jordi Cabot}
\IEEEauthorblockA{
\textit{Luxembourg Institute of Science and Technology}\\
\textit{University of Luxembourg}\\
Esch-sur-Alzette, Luxembourg \\
0000-0003-2418-2489}
}


\maketitle

\begin{abstract}
Multimodal systems, which process multiple input types such as text, audio, and images, are becoming increasingly prevalent in software systems, enabled by the huge advancements in Machine Learning. This triggers the need to easily define the requirements linked to these new types of user interactions, potentially involving more than one modality at the same time. This remains an open challenge due to the lack of languages and methods adapted to the diverse nature of multimodal interactions, with the risk of implementing AI-enhanced systems that do not properly satisfy the user needs.

In this sense, this paper presents MERLAN, a Domain-Specific Language (DSL) to specify the requirements for these new types of multimodal interfaces. We present the metamodel for such language together with a textual syntax implemented as an ANTLR grammar. A prototype tool enabling requirements engineers to write such requirements and automatically generate a possible implementation of a system compliant with them on top of an agentic framework is also provided.




\end{abstract}


\input{sections/01_introduction}

\input{sections/02_running_example}
\input{sections/03_abstract_syntax}
\input{sections/04_concrete_syntax}
\input{sections/05_tool_support}

\input{sections/06_related_work}

\input{sections/07_research_roadmap}

\input{sections/08_conclusions}

\bibliographystyle{IEEEtran}
\bibliography{references.bib}

\end{document}

%% file: sections/01_introduction.tex
\section{Introduction}
\label{sec:introduction}

With the explosion of Machine Learning and other AI techniques, software systems are quickly adopting new types of complex user interfaces that require processing new input modalities such as text, audio and images. Sometimes, more than one type at the same time. This type of interfaces are known as Multimodal User Interfaces (MUIs). MUIs provide an interface that bears the functionalities of human-human interface \cite{mui}. It is a broad field within Human-Computer Interaction (HCI) with different applications and definitions depending on the context. The term ``modality'' refers to the mode of communication used as input and output between the human and the computer, like text, audio and image, although other modalities are considered in other contexts (e.g., facial expression, gestures, etc.).

 

While building this type of AI-enhanced systems is becoming easier thanks to the constant influx of, for instance, new multimodal Large Language Models (LLMs) that facilitate the analysis of multimodal inputs, validating that the system is satisfying the actual needs of the user is becoming more and more complex. Indeed, we are missing proper requirements engineering languages and techniques to facilitate a precise specification of the MUI conditions that should trigger a system response, the data (``entities'' in the MUI terminology) from the multimodal input that should be collected to provide an adequate response and the actual response \cite{guidelines}. 

For the latter, current languages could be reused, but a new way to specify MUI-driven requirements is needed for the first two elements. Preliminary work in this area has focused on requirements for chatbots. In particular, to identify what user ``intents'' (i.e. expressions of users' intentions or goals) the chatbot should respond to, and what entities or parameters are part of the user textual input matching to a specific intent \cite{xatkit} . However, this is not the case with other modalities (e.g., image or audio) and even less for requirements involving multiple inputs.
 



To solve this problem, this paper presents MERLAN: a Multimodal Environment Requirements Language\footnote{\url{https://github.com/BESSER-PEARL/merlan}}. MERLAN aims to standardize the definition of multimodal requirements, independently of the underlying technology being used to process multimodal data. MERLAN is a Domain-Specific Language (DSL) formalized with a language metamodel and supported by a textual concrete syntax to help the users easily define MUI requirements. Tool support is also available. It consist of the language support but also a proof-of-concept implementation of a transformation from the requirements to a Python-based agent able to trigger system responses when the requirements conditions are met.


The rest of the paper is structured as follows: Section \ref{sec:running-example} introduces a running example to motivate and illustrate our DSL; Section \ref{sec:abstract-syntax} presents the abstract syntax of the DSL while Section \ref{sec:concrete-syntax} focuses on its concrete syntax; Section \ref{sec:tool-support} shows the provided tool support; Section \ref{sec:related-work} discusses the related work of this paper; The research roadmap is discussed in Section \ref{sec:research-roadmap} and, finally, Section \ref{sec:conclusions} closes with the conclusions.

%% file: sections/02_running_example.tex
\section{Running Example}
\label{sec:running-example}

To motivate and illustrate our approach, this section introduces a running example that will be used throughout the paper. 

Let's imagine that we need to specify the requirements for a new home automation and security system, hereinafter referred to as house agent. The house agent has some input devices to capture text, sound, video, temperature, light and movement. It also has output devices for text and audio, in addition to the capability of executing a set of predefined actions  (i.e., actions such as making a call or triggering the alarm).

Regardless of the actual software components in charge of capturing information from the physical world (e.g., ML models to detect objects from camera inputs), the house agent needs to be provided with clear definitions of the entities that can be recognized from all the different input devices, and the rules that describe how those entities should be evaluated by the agent in order to decide whether to trigger a certain response. 

For instance, regarding the image input devices, we could define concrete entities such as \verb|person|, \verb|dog|, \verb|car|, \verb|smoke| and \verb|fire|. An audio entity could be \verb|strong_sound|. Moreover, these entities can have attributes. The \verb|person| entity could have \verb|gender| and \verb|ethnicity| attributes, which during the recognition process of the agent, should be filled with the right values. The system needs other kind of more abstract entities to be defined, such as ``day or night'' or ``empty house''. They are more abstract in the sense that they must be  inferred from parts of the input data instead of being directly identified as objects in it. 

With all these entities defined, we can design the rules the agent will check to trigger some actions. For instance:

\begin{enumerate}
  \item If smoke is detected: notify the house owner.
  \item If (fire is detected) or (the house is empty and (some cars or some persons are detected)): Activate the alarm, notify the house owner and call the police.
  \item If (strong sound) and (during night): turn on lights
  \item If a car with unrecognized plate number is detected: notify the house owner
\end{enumerate}

The first rule simply defines the presence of a concrete entity (smoke). The second consists of the composition of multiple concrete and abstract requirements with ``and'' and ``or'' operators. The third expects the detection of an audio concrete entity (strong sound) and an image abstract entity (night). The last requirement is satisfied when not only an entity (car) but also a specific attribute value (the license plate) are identified. 

This set of rules and entities  would compose the requirements of the house agent's multimodal interface, describing what conditions and based on what data the system should react to, and the actions to be taken in case of requirements satisfaction. These requirements could be described using natural language, but without enough precision to implement them with confidence. 

Therefore, we believe a DSL to enable the formal definition of MUI requirements in a more precise way is needed.  The next section introduces MERLAN, the DSL we propose for this.  


%% file: sections/03_abstract_syntax.tex
\section{DSL Design}
\label{sec:abstract-syntax}

A Domain-Specific Language (DSL) is a specialized programming or modeling language designed to address problems within a specific domain. Unlike general-purpose languages, DSLs provide higher-level abstractions tailored to the concepts and rules of the domain, making it easier to express models, transformations, and constraints \cite{dsls}. A DSL is defined by an abstract syntax that specifies the DSL main concepts (and their interrelationships) and a concrete syntax that implements it (usually via a textual or graphical notation).

The abstract syntax of a DSL defines its structural representation via a metamodel specification independent of  any concrete syntax or notation. This section presents the metamodel of the core abstract syntax of MERLAN for expressing MUI requirements, outlining its key elements and relationships. 

\begin{figure*}[ht!]
    \centering
    \includegraphics[width=\linewidth]{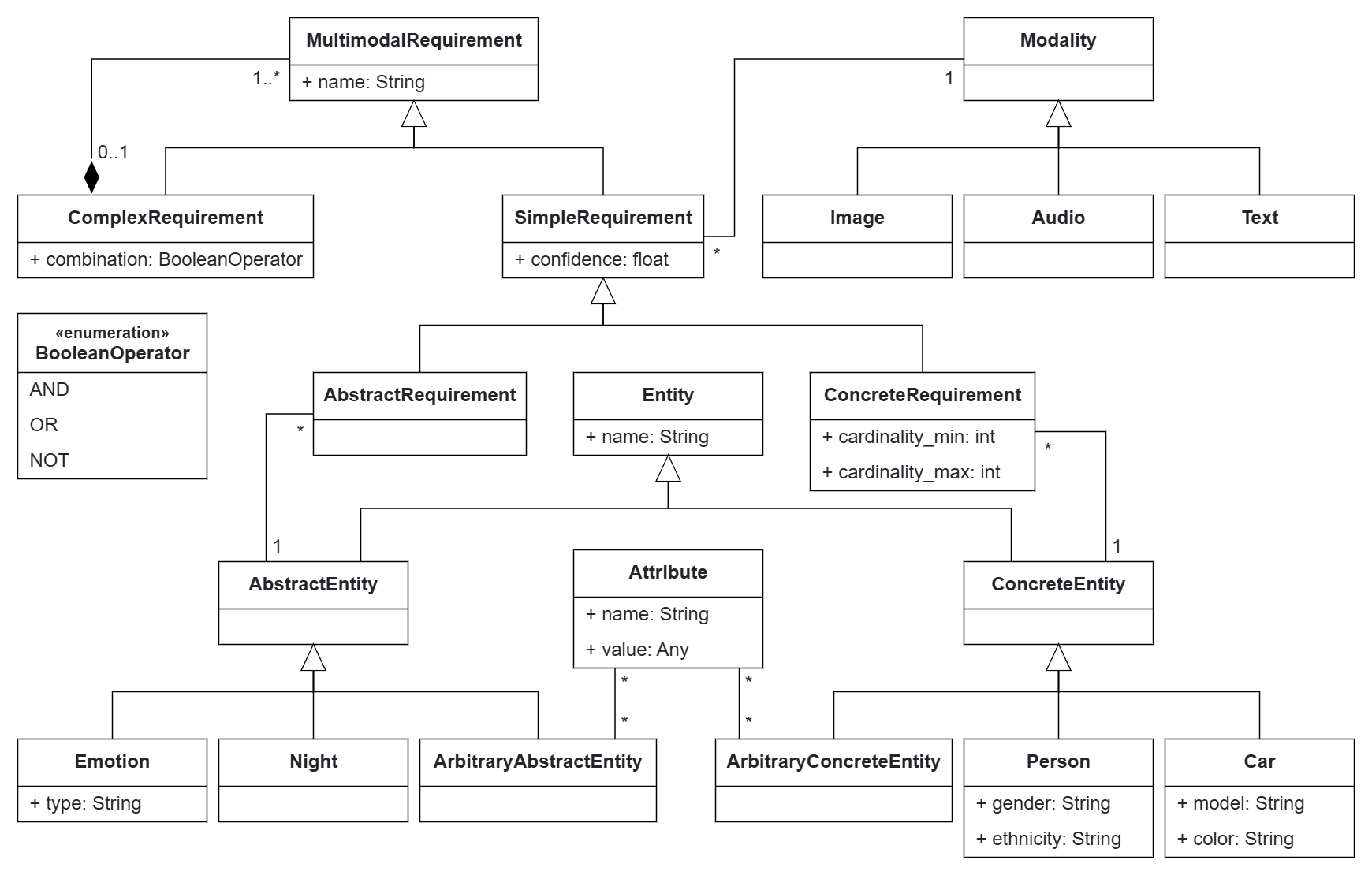}
    \caption{MERLAN Metamodel}
    \label{fig:metamodel}
\end{figure*}

Figure \ref{fig:metamodel} illustrates this metamodel expressed using a UML class diagram formalism, as usual. 

Next subsections describe in more detail the main elements of the metamodel. We split the explanation in two subections, one covering the metaclasses focused on the definition of Multimodal Requirements and one focusing more on the definition of Entities referenced in those requirements.

\subsection{Definition of Multimodal Requirements}

A \verb|MultimodalRequirement| defines the rules that need to be evaluated in a system with multimodal inputs. These rules provide formal definitions of conditions the system should match in order to trigger some actions. 

The metamodel defines two kinds of requirements: simple and complex.

    \subsubsection{Complex Requirements}

    A \verb|ComplexRequirement| is a composition of \verb|MultimodalRequirements|, which can be themselves also complex or simple. Requirements are composed with boolean operators, namely \verb|AND| (all requirements under it must be satisfied), \verb|OR| (at least one requirement must be satisfied) and \verb|NOT| (the requirement condition under it must not be satisfied).

    \subsubsection{Simple Requirements}

    A \verb|SimpleRequirement| expresses a condition on a single entity. When defining a simple requirement, we are creating a rule that would be satisfied when an entity is detected. Simple requirements have a \verb|confidence| attribute to indicate the minimum confidence that needs to be achieved for the requirement to be satisfied (i.e., confidence on the entity detection).

    \verb|SimpleRequirement| has \verb|AbstractRequirement| and \verb|ConcreteRequirement| to distinguish between simple requirements referencing concrete and abstract entities, respectively. The difference between them is that a \verb|ConcreteRequirement| can define a cardinality to specific the number of instances of that entity that should be detected. The cardinality notation is based on the UML cardinalities, allowing to define specific values (\verb|[n]|, exactly n instances), concrete intervals (\verb|[m..n]|, where $m<n$ and $m>=0$) or unbounded intervals (\verb|[n..*]|, where $n>=0$). As we will see later, entities can have attributes which can also be used during the matching process. 
    
    
    A \verb|SimpleRequirement| has a specific modality, indicating which modality should be considered to satisfy the requirement. For instance, a \verb|SimpleRequirement| referencing a \verb|Person| concrete entity with \verb|Image| modality will be evaluated based on image inputs (i.e., looking for a person in an image), while the same requirement with \verb|Audio| modality will analyze the audio input to find a person (i.e., if someone is speaking and the voice is detected). When evaluating the requirements, some attributes make sense only for specific modalities (e.g., the \verb|color| of a \verb|Car| entity only makes sense for image modality). This is something to consider when defining the entity attributes.

\subsection{Definition of Entities}

An \verb|Entity| represents a real-world concept, either material or abstract, that can be distinctly identified in some type of multimodal input. The simplest way to define entities is via a name and attributes that specify its structure and content. When defining an entity, we define its attributes, which may have specific values or empty ones (i.e., unknown values). 

Setting an attribute value enforces the recognition engine to identify those entities with that exact attribute value. For instance, if we define a \verb|Person| entity with \verb|gender = "male"| attribute, we could use that entity within a requirement to restrict its applicability to men.

An attribute value can be left empty with the purpose of dynamically inferring its value during the recognition process. Following the \verb|Person| entity example, a \verb|gender| attribute with empty value should make the recognition engine identify both men and women and fill the \verb|gender| value of the recognized \verb|Person| at runtime. Attributes with empty values also serve as placeholders that can be set at the requirement level. This mechanism provides flexibility by allowing the definition of generic entities and deferring context-specific details to individual requirements.

We consider two types of entities based on their nature, namely abstract and concrete entities.

    \subsubsection{Concrete Entities}
    
    A \verb|ConcreteEntity| is a subclass of \verb|Entity| that represents a concrete object or being within the multimodal environment that can be described in terms of its physical existence and referenced or interacted within the environment.

    \subsubsection{Abstract Entities}
    
    \verb|AbstractEntity| is a subclass of \verb|Entity| that does not have a clearly bounded physical existence but represents a concept, idea, property or classification that can be inferred from the environment. An example abstract entity could be \verb|night|, indicating if an image is taken at night or not. A more abstract one could be \verb|hazard|, indicating the level of hazard perceived from an image or audio (we could define some attributes for this property to describe our ``hazard criteria'').
    
    \subsubsection{Predefined and Arbitrary Entities}
    
    The metamodel figure includes example instantiations of both concrete and abstract entities with default attributes. For instance, a simplified \verb|Person| concrete entity can be defined in terms of the attributes \verb|gender| and \verb|ethnicity|. Abstract and concrete entities have an \verb|ArbitraryEntity| subclass that allows the definition of custom or domain-specific entities with arbitrary attributes. 

    To simplify the definition of requirements in a given domain we could predefine the most relevant entities in that domain based on existing ontologies for the domain. This is a trade-off between the size of the DSL and its ease of use. 

%% file: sections/04_concrete_syntax.tex
\definecolor{customgray}{rgb}{0.95,0.95,0.95}  
\definecolor{customgreen}{rgb}{0.0, 0.5, 0.0}  

\lstdefinelanguage{ANTLR}{
    morekeywords={MERLAN, NEWLINE, ENTITIES, CONCRETE, ID, AND, OR, NOT},
    sensitive=true,
    morecomment=[l]{//},
    morecomment=[s]{/*}{*/},
    morestring=[b]{"}
}

\lstdefinelanguage{MERLAN}{
    morekeywords={ENTITIES, CONCRETE, ABSTRACT, REQUIREMENTS, AND, OR, NOT, IMAGE_PROPERTY, IMAGE_ENTITY},
    sensitive=true,
    morecomment=[l]{//},
    morecomment=[s]{/*}{*/},
    morestring=[b]{"}
}

\lstdefinestyle{ANTLRStyle}{
    language=ANTLR,
    backgroundcolor=\color{customgray},
    basicstyle=\ttfamily\small,
    keywordstyle=\color{blue},
    commentstyle=\itshape\color{gray},
    stringstyle=\color{customgreen},
    numbers=left,
    numbersep=2pt,
    numberstyle=\tiny\color{gray},
    showstringspaces=false,
    frame=none,
    breaklines=true,  
    breakatwhitespace=true,  
    xleftmargin=6pt  
}

\lstdefinestyle{MERLANStyle}{
    language=MERLAN,
    backgroundcolor=\color{customgray},
    basicstyle=\ttfamily\small,
    keywordstyle=\color{blue},
    commentstyle=\itshape\color{gray},
    stringstyle=\color{customgreen},
    numbers=left,
    numbersep=2pt,
    numberstyle=\tiny\color{gray},
    showstringspaces=false,
    frame=none,
    breaklines=true,  
    breakatwhitespace=true,  
    xleftmargin=6pt  
}

\section{Concrete Syntax}
\label{sec:concrete-syntax}


This section introduces the concrete syntax of our DSL, which conforms to the previously detailed metamodel. The concrete syntax defines the rules ensuring a consistent and standardized use of the language. In this paper, we present a textual concrete syntax implemented with ANTLR, a powerful tool to generate language parsers from grammars. 

We start by introducing the grammar and next an example MERLAN specification following the grammar. Listing \ref{lst:grammar-1} presents an excerpt of the DSL grammar.

\begin{lstlisting}[style=ANTLRStyle, caption={MERLAN ANTLR grammar; main rules.}, label={lst:grammar-1}]
grammar MERLAN;

script
  : entities?
    requirements?
  ;

entities
  : ENTITIES NEWLINE
    concrete_entities?
    abstract_entities?
  ;

concrete_entities
  : CONCRETE NEWLINE concrete_entity*
  ;

concrete_entity
  : ID NEWLINE attribute*
  ;

// Other rules omitted for brevity

requirement
  : complex_requirement
  | simple_requirement
  ;

complex_requirement
  : (AND | OR) NEWLINE requirement+
  | NOT NEWLINE requirement
  ;

simple_requirement
  : abstract_requirement
  | concrete_requirement
  ;

concrete_requirement
  : CONCRETE cardinality? NEWLINE attribute*
  ;
  
// Other rules omitted for brevity
\end{lstlisting}

The grammar defines the syntactic rules allowed in MERLAN language. At the first level, the \verb|script| rule indicates that \verb|entities| and \verb|requirements| can be defined.

Some rules start with an uppercase keyword to divide the code into clear sections (e.g., see \verb|ENTITIES|). The \verb|entities| section can include concrete or abstract entities. For brevity, we excluded the \verb|abstract_entities| rule (among others) in this paper, but it is analogous to \verb|concrete_entities|. A concrete entity is defined by its name (i.e., the \verb|ID|) and any number of attributes.

A \verb|requirement| can be either complex or simple. A \verb|complex_requirement| starts with a boolean operator (\verb|AND|, \verb|OR| or \verb|NOT|) and a set of requirements (limited to 1 if the operator is \verb|NOT|). A \verb|simple_requirement| can be either abstract or concrete, where \verb|concrete_requirement| can have a custom cardinality. Simple requirements contain a set of attributes, where some of them are mandatory. Mandatory attributes are not specified since the grammar's purpose is to ensure syntactical consistency. Evaluation on the missing attributes (as well as cardinality correctness) is done during the parser execution (i.e., when parsing a MERLAN code script). Mandatory attributes include \verb|confidence|, \verb|modality| and \verb|entity| (those apply to simple requirements only) and \verb|name|.


Listing \ref{lst:merlan-example} shows an example MERLAN code that defines entities and requirements following the previously proposed running example of a house agent in Section \ref{sec:running-example}. 

\begin{lstlisting}[style=MERLANStyle, caption={Example MERLAN code.}, label={lst:merlan-example}]
ENTITIES:
  CONCRETE:
    person
      - gender: ?
      - ethnicity: ?
    smoke
    fire
    dog
      - breed: "labrador"
    car:
      - model: ?
      - color: ?
  ABSTRACT:
    night
      - description: "The image is taken at night"
    empty_house
      - description: "The house is empty"
REQUIREMENTS:
  requirement1:
    CONCRETE
      - entity: smoke
      - name: "smoke"
      - modality: "image"
      - confidence: 0.5
  requirement2
    OR
      CONCRETE
      - entity: fire
      - name: "fire"
      - confidence: 0.5
      - modality: "image"
      AND
        ABSTRACT
          - entity: empty_house
          - name: "empty_house"
          - confidence: 0.3
          - modality: "image"
        OR
          CONCRETE [1..*]
            - entity: person
            - name: "unknown_person"
            - confidence: 0.7
            - modality: "image"
            - gender: "male"
          CONCRETE [1..*]
            - entity: car
            - name: "unknown_car"
            - confidence: 0.7
            - modality: "image"
\end{lstlisting}

The first code block, identified with the \verb|ENTITIES| keyword, contains all the entity definitions following the grammar rules. In this example, there are entities without attributes (see \verb|smoke| and \verb|fire|), an entity \verb|dog| with an attribute with specific value (\verb|breed: "labrador"|) and other entities (\verb|person| and \verb|car|) with attributes with empty values.

The requirements block, under the \verb|REQUIREMENTS| keyword, contains the 2 example requirements defined in Section \ref{sec:running-example}. \verb|requirement2| contains a composition of complex requirements where 2 of the inner simple requirements define cardinalities of \verb|[1..*]| (i.e., minimum 1 instance). The requirement referencing the \verb|person| entity shows how to set an entity attribute's value at the requirement level (see the \verb|gender: "male"| attribute).

%% file: sections/05_tool_support.tex
\section{Tool Support}
\label{sec:tool-support}

The full ANTLR grammar to use MERLAN is available in our open source repository in GitHub.

With the grammar, requirement engineers can describe full MERLAN specifications. But to make these specifications more actionable and offer a better Return On Investment (ROI) from them, we have also implemented a prproof-of-concept implementation to derive an agent implementation that understand and reacts to the requirement conditions effectively triggering a change on state based on detecting the satisfiability of the conditions on a multimodel input.

More specifically, we have implemented a transformation that, given a MERLAN specification creates an agent implemented on top of the BESSER Agentic Framework\footnote{\url{https://github.com/BESSER-PEARL/BESSER-Agentic-Framework}} where the MERLAN requirements have been transformed into a set of agent triggering conditions.  Listing \ref{lst:python-example} shows the generated code for the example MERLAN code in Listing \ref{lst:merlan-example}

The transformation is implemented via a custom implementation of a tree visitor pattern that traverses the abstract syntax tree generated by the MERLAN ANTLR parser. Note that this approach can be eaisly replicated for any language or agentic framework that supports similar requirements-based transitions.

The agentic framework we selected  comes with integrated LLMs and Computer Vision models that capture information from a multimodal environment, which is used in real time to evaluate the requirements matching. 
The generated code includes multimodal entities, attached to the agent's entity database, and the requirements' conditions themselves. Based on this input processing component,  the agent developer can complete the agent specification expressing the response behaviour to be executed in response of a triggering condition. 

\lstdefinestyle{PythonStyle}{
    language=Python,
    backgroundcolor=\color{customgray},
    basicstyle=\ttfamily\small,
    keywordstyle=\color{blue},
    commentstyle=\itshape\color{gray},
    stringstyle=\color{customgreen},
    numbers=left,
    numbersep=2pt,
    numberstyle=\tiny\color{gray},
    showstringspaces=false,
    frame=none,
    breaklines=true,  
    breakatwhitespace=true,  
    xleftmargin=6pt  
}

\begin{figure*}[t]
    \centering
    \begin{minipage}{\textwidth}
        \begin{lstlisting}[style=PythonStyle, caption={Generated Python code.}, label={lst:python-example}]
# Entities
person = ConcreteEntity(name="person", attributes={"gender": None, "ethnicity": None})
smoke = ConcreteEntity(name="smoke", attributes={})
fire = ConcreteEntity(name="fire", attributes={})
dog = ConcreteEntity(name="dog", attributes={"breed": "labrador"})
car = ConcreteEntity(name="car", attributes={"brand": None, "model": None, "color": None})
night = AbstractEntity(name="night", attributes={"description": "The image is taken at night"})
empty_house = AbstractEntity(name="empty_house", attributes={"description": "The house is empty"})
# Requirements
requirement1 = RequirementDefinition("requirement1")
requirement1.set(ConcreteRequirement(name="smoke", concrete_entity=smoke, attributes={"modality": "image", "confidence": 0.5}))
requirement2 = RequirementDefinition("requirement2")
requirement2.set(
  OR([
    ConcreteRequirement(name="fire", concrete_entity=fire, attributes={"confidence": 0.5, "modality": "image"}),
    AND([
      AbstractRequirement(name="empty_house", abstract_entity=empty_house, attributes={"confidence": 0.3, "modality": "image"}),
      OR([
        ConcreteRequirement(name="man", concrete_entity=person, attributes={"min": 1, "max": 0, "confidence": 0.7, "modality": "image", "gender": "male"}),
        ConcreteRequirement(name="unknown_car", concrete_entity=car, attributes={"min": 1, "max": 0, "confidence": 0.7, "modality": "image"})
      ])
    ])
  ])
)

# The following code is not automatically generated. It ilustrates how to use a requirement to define transitions between the agent states (agent definition code ommited for brevity)
initial_state.when_requirement_matched_go_to(requirement1, smoke_state)
        \end{lstlisting}
    \end{minipage}
\end{figure*}

%% file: sections/06_related_work.tex
\section{Related Work}
\label{sec:related-work}

In this section, we discuss the work related to the specification of requirements for MUIs. We focus first on frameworks that explicitly address the creation of MUIs to then cover other approaches closer to the requirements engineering field, proposing languages to express requirements for some types of advanced interfaces. 


\subsection{Multimodal User Interfaces}



Recent work around MUIs is focusing on the exploitation of  Machine Learning techniques to automatically extract patterns and insights from data \cite{multimodal-ml}. 

A variety of frameworks have been proposed for this purpose. 
Xspeak \cite{xspeak} added a speech interface on top of X Window System, allowing to use words to interact with windows. Open Agent Architecture (OAA) \cite{oaa} proposed a framework for multiagent systems with MUIs that included spoken language, handwriting and gesture. Openinterface \cite{openinterface} is another tool to design MUIs, providing its own runtime and IDE. The Squidy Interaction Library \cite{Konig2010} proposed a library to reduce the efforts of designing MUIs, integrating different toolkits and frameworks in a common library integrating a GUI, hiding complexity by providing a visual language and a collection of devices and interaction techniques.

There is also a particular interest on MUIs in the robotics and healthcare domains. For instance, AMIR \cite{electronics9122093} is an assistive robot with voice and gesture-based interfaces, and FIRMA \cite{firma}, a development framework for elderly-friendly interactive multimodal applications for assistive robots. MUIs also have a strong presence in smart home user interfaces \cite{aal}.

Nevertheless, all these approaches focus on the development of the MUIs themselves not on the formalization of the requirements they are supposed to implement, which is exactly the purpose of our MERLAN proposal. 


\subsection{Domain-Specific Languages}

DSLs are used in many Machine Learning problems \cite{ml-dsls}, including DSLs that focus on formalizing requirements \cite{9920140}. An example is Impromptu \cite{Morales2025} proposed a DSL to define structured prompts in a modular and tool-independent way. Other DSLs focus more on fairness aspects of Machine Learning, e.g. \cite{KolovosFairML, GinerDescribeML}.

Closer to our work, other DSLs cover multimodal aspects of software components. For instance, SEMKIS \cite{SEMKIS} focus on requirements engineering of datasets and neural networks to improve recognition skills.  ViSaL \cite{visal} allows the programmer to express image quality detection rules for enforcing safety constraints, increasing trustworthiness in robot perception systems. FVision \cite{fvision} designed as a Haskell library as a DSL to build and test visual tracking systems. The Midgar IoT platform was extended by adding a Computer Vision module to automate camera input analysis. Their approach detects only people, requiring model training for other objects. They propose developing a DSL to streamline the Computer Vision pipeline \cite{Midgar}.  Note that there are existing resources to evaluate such systems, such as CLEVR \cite{clevr}, which proposed a dataset to benchmark visual reasoning in smart systems. More specific to the chatbot domain, \cite{xatkit} and \cite{deLara20} propose DSLs for intent matching and entity recognition in textual inputs. 

Note that the above examples target specific types of inputs and, many of them, specific environments, while MERLAN aims to provide a multimodal requirements solution combining different types of requirements and entities (including concrete and abstract ones) and conditions on them, opening the door to the creation of powerful multimodal agents that satisfy the user requirements.

%% file: sections/07_research_roadmap.tex
\section{Research Roadmap}
\label{sec:research-roadmap}

To fully define a robust and expressive Domain-Specific Requirements Language for Multimodal User Interfaces, several key aspects must be addressed. In what follows we list some of them.
    
\begin{itemize}

    \item Add a graphical notation to MERLAN, enhancing usability and making it easier for less-technical practitioners and engineers to specify and visualize MUI requirements. This should even include a visual modeling by example component where users could give images as example scenarios that should trigger an action in the system. Alternative, also the option to simply describe a requirement in natural language. Machine Learning can be used to support this task.
    
    \item Extend the language to cover also the specification of requirements on the behaviour to perform when the conditions are matched. For "traditional" actions, existing behavioural languages (e.g. UML specifications) could suffice but for multimodal responses, an extension to MERLAN, where multimodality is already a first-class element could be a better option.

    \item Time conditions to enable expressing temporal constraints in the requirements, e.g.. constraints on the duration of a certain object in a video before triggering an action 

    \item Additionally, hierarchical modalities should be explored, where higher-level modalities (e.g., gestures or facial expressions) are derived from lower-level ones (e.g., images or video). This hierarchical structuring will enable more precise and flexible requirements definitions for complex multimodal interactions. 
    
    \item Quality analysis to detect inconsistencies and conflicts among multimodal requirements. For example, certain attributes, such as color, are relevant only in image-based modalities and may not be applicable in textual or auditory contexts. Or more challenging to detect, two conditions in an image may be mutually exclusive, implying that such condition can never be satisfied.
    
\end{itemize}


%% file: sections/08_conclusions.tex
\section{Conclusions}
\label{sec:conclusions}

In this paper, we introduced MERLAN, a Domain-Specific Language designed to standardize the specification of requirements for Multimodal User Interfaces (MUIs). 
Our approach leverages a metamodel-based formalization and a textual concrete syntax to facilitate the specification process while also providing a proof-of-concept implementation showing a potential path to transform these requirements into actionable agent code to execute the requirements in real-time multimodal environments.

Future work will focus on expanding the capabilities of MERLAN by addressing the open challenges discussed above. 
Additionally,  we aim to conduct empirical validation experiments and extend our tool support to foster adoption our language and infrastructure by researchers and practitioners in the field.  As part of such tool extension, we plan to develop a library of predefined entities, based on existing ontologies, to be imported and reused when expressing new requirements.

%% file: main.bbl
\begin{thebibliography}{10}
\providecommand{\url}[1]{#1}
\csname url@samestyle\endcsname
\providecommand{\newblock}{\relax}
\providecommand{\bibinfo}[2]{#2}
\providecommand{\BIBentrySTDinterwordspacing}{\spaceskip=0pt\relax}
\providecommand{\BIBentryALTinterwordstretchfactor}{4}
\providecommand{\BIBentryALTinterwordspacing}{\spaceskip=\fontdimen2\font plus
\BIBentryALTinterwordstretchfactor\fontdimen3\font minus
  \fontdimen4\font\relax}
\providecommand{\BIBforeignlanguage}[2]{{%
\expandafter\ifx\csname l@#1\endcsname\relax
\typeout{** WARNING: IEEEtran.bst: No hyphenation pattern has been}%
\typeout{** loaded for the language `#1'. Using the pattern for}%
\typeout{** the default language instead.}%
\else
\language=\csname l@#1\endcsname
\fi
#2}}
\providecommand{\BIBdecl}{\relax}
\BIBdecl

\bibitem{mui}
\BIBentryALTinterwordspacing
M.~Z. Baig and M.~Kavakli, ``Multimodal systems: Taxonomy, methods, and
  challenges,'' 2020. [Online]. Available:
  \url{https://arxiv.org/abs/2006.03813}
\BIBentrySTDinterwordspacing

\bibitem{guidelines}
\BIBentryALTinterwordspacing
L.~M. Reeves, J.~Lai, J.~A. Larson, S.~Oviatt, T.~S. Balaji, S.~Buisine,
  P.~Collings, P.~Cohen, B.~Kraal, J.-C. Martin, M.~McTear, T.~Raman, K.~M.
  Stanney, H.~Su, and Q.~Y. Wang, ``Guidelines for multimodal user interface
  design,'' \emph{Commun. ACM}, vol.~47, no.~1, p. 57–59, Jan. 2004.
  [Online]. Available: \url{https://doi-org.proxy.bnl.lu/10.1145/962081.962106}
\BIBentrySTDinterwordspacing

\bibitem{xatkit}
G.~Daniel, J.~Cabot, L.~Deruelle, and M.~Derras, ``Xatkit: A multimodal
  low-code chatbot development framework,'' \emph{IEEE Access}, vol.~8, pp.
  15\,332--15\,346, 2020.

\bibitem{dsls}
\BIBentryALTinterwordspacing
M.~Mernik, J.~Heering, and A.~M. Sloane, ``When and how to develop
  domain-specific languages,'' \emph{ACM Comput. Surv.}, vol.~37, no.~4, p.
  316–344, Dec. 2005. [Online]. Available:
  \url{https://doi-org.proxy.bnl.lu/10.1145/1118890.1118892}
\BIBentrySTDinterwordspacing

\bibitem{multimodal-ml}
T.~Baltrušaitis, C.~Ahuja, and L.-P. Morency, ``Multimodal machine learning: A
  survey and taxonomy,'' \emph{IEEE Transactions on Pattern Analysis and
  Machine Intelligence}, vol.~41, no.~2, pp. 423--443, 2019.

\bibitem{xspeak}
C.~Schmandt, M.~Ackerman, and D.~Hindus, ``Augmenting a window system with
  speech input,'' \emph{Computer}, vol.~23, no.~8, pp. 50--56, 1990.

\bibitem{oaa}
\BIBentryALTinterwordspacing
D.~B. Moran, A.~J. Cheyer, L.~E. Julia, D.~L. Martin, and S.~Park, ``Multimodal
  user interfaces in the open agent architecture,'' in \emph{Proceedings of the
  2nd International Conference on Intelligent User Interfaces}, ser. IUI
  '97.\hskip 1em plus 0.5em minus 0.4em\relax New York, NY, USA: Association
  for Computing Machinery, 1997, p. 61–68. [Online]. Available:
  \url{https://doi-org.proxy.bnl.lu/10.1145/238218.238290}
\BIBentrySTDinterwordspacing

\bibitem{openinterface}
\BIBentryALTinterwordspacing
M.~Serrano, L.~Nigay, J.-Y.~L. Lawson, A.~Ramsay, R.~Murray-Smith, and
  S.~Denef, ``The openinterface framework: a tool for multimodal interaction.''
  in \emph{CHI '08 Extended Abstracts on Human Factors in Computing Systems},
  ser. CHI EA '08.\hskip 1em plus 0.5em minus 0.4em\relax New York, NY, USA:
  Association for Computing Machinery, 2008, p. 3501–3506. [Online].
  Available: \url{https://doi.org/10.1145/1358628.1358881}
\BIBentrySTDinterwordspacing

\bibitem{Konig2010}
\BIBentryALTinterwordspacing
W.~A. K{\"o}nig, R.~R{\"a}dle, and H.~Reiterer, ``Interactive design of
  multimodal user interfaces,'' \emph{Journal on Multimodal User Interfaces},
  vol.~3, no.~3, pp. 197--213, Apr 2010. [Online]. Available:
  \url{https://doi.org/10.1007/s12193-010-0044-2}
\BIBentrySTDinterwordspacing

\bibitem{electronics9122093}
\BIBentryALTinterwordspacing
D.~Ryumin, I.~Kagirov, A.~Axyonov, N.~Pavlyuk, A.~Saveliev, I.~Kipyatkova,
  M.~Zelezny, I.~Mporas, and A.~Karpov, ``A multimodal user interface for an
  assistive robotic shopping cart,'' \emph{Electronics}, vol.~9, no.~12, 2020.
  [Online]. Available: \url{https://www.mdpi.com/2079-9292/9/12/2093}
\BIBentrySTDinterwordspacing

\bibitem{firma}
N.~Kazepis, M.~Antona, and C.~Stephanidis, ``Firma: A development framework for
  elderly-friendly interactive multimodal applications for assistive robots,''
  04 2016.

\bibitem{aal}
M.~Blumendorf and S.~Albayrak, ``Towards a framework for the development of
  adaptive multimodal user interfaces for ambient assisted living
  environments,'' in \emph{Universal Access in Human-Computer Interaction.
  Intelligent and Ubiquitous Interaction Environments}, C.~Stephanidis,
  Ed.\hskip 1em plus 0.5em minus 0.4em\relax Berlin, Heidelberg: Springer
  Berlin Heidelberg, 2009, pp. 150--159.

\bibitem{ml-dsls}
I.~Portugal, P.~Alencar, and D.~Cowan, ``A preliminary survey on
  domain-specific languages for machine learning in big data,'' in \emph{2016
  IEEE International Conference on Software Science, Technology and Engineering
  (SWSTE)}, 2016, pp. 108--110.

\bibitem{9920140}
Z.~Pei, L.~Liu, C.~Wang, and J.~Wang, ``Requirements engineering for machine
  learning: A review and reflection,'' in \emph{2022 IEEE 30th International
  Requirements Engineering Conference Workshops (REW)}, 2022, pp. 166--175.

\bibitem{Morales2025}
\BIBentryALTinterwordspacing
S.~Morales, R.~Claris{\'o}, and J.~Cabot, ``Impromptu: a framework for
  model-driven prompt engineering,'' \emph{Software and Systems Modeling}, Jan
  2025. [Online]. Available: \url{https://doi.org/10.1007/s10270-024-01235-4}
\BIBentrySTDinterwordspacing

\bibitem{KolovosFairML}
\BIBentryALTinterwordspacing
A.~Yohannis and D.~Kolovos, ``Towards model-based bias mitigation in machine
  learning,'' in \emph{Proceedings of the 25th International Conference on
  Model Driven Engineering Languages and Systems}, ser. MODELS '22.\hskip 1em
  plus 0.5em minus 0.4em\relax New York, NY, USA: Association for Computing
  Machinery, 2022, p. 143–153. [Online]. Available:
  \url{https://doi.org/10.1145/3550355.3552401}
\BIBentrySTDinterwordspacing

\bibitem{GinerDescribeML}
\BIBentryALTinterwordspacing
J.~Giner{-}Miguelez, A.~G{\'{o}}mez, and J.~Cabot, ``Describeml: {A} dataset
  description tool for machine learning,'' \emph{Sci. Comput. Program.}, vol.
  231, p. 103030, 2024. [Online]. Available:
  \url{https://doi.org/10.1016/j.scico.2023.103030}
\BIBentrySTDinterwordspacing

\bibitem{SEMKIS}
\BIBentryALTinterwordspacing
B.~Jahić, N.~Guelfi, and B.~Ries, ``Semkis-dsl: A domain-specific language to
  support requirements engineering of datasets and neural network
  recognition,'' \emph{Information}, vol.~14, no.~4, 2023. [Online]. Available:
  \url{https://www.mdpi.com/2078-2489/14/4/213}
\BIBentrySTDinterwordspacing

\bibitem{visal}
J.~T.~M. Ingibergsson, D.~Kraft, and U.~P. Schultz, ``Safety computer vision
  rules for improved sensor certification,'' in \emph{2017 First IEEE
  International Conference on Robotic Computing (IRC)}, 2017, pp. 89--92.

\bibitem{fvision}
J.~Peterson, P.~Hudak, A.~Reid, and G.~Hager, ``Fvision: A declarative language
  for visual tracking,'' in \emph{Practical Aspects of Declarative Languages},
  I.~V. Ramakrishnan, Ed.\hskip 1em plus 0.5em minus 0.4em\relax Berlin,
  Heidelberg: Springer Berlin Heidelberg, 2001, pp. 304--321.

\bibitem{Midgar}
\BIBentryALTinterwordspacing
C.~{González García}, D.~Meana-Llorián, B.~C. {Pelayo G-Bustelo}, J.~M.
  {Cueva Lovelle}, and N.~Garcia-Fernandez, ``Midgar: Detection of people
  through computer vision in the internet of things scenarios to improve the
  security in smart cities, smart towns, and smart homes,'' \emph{Future
  Generation Computer Systems}, vol.~76, pp. 301--313, 2017. [Online].
  Available:
  \url{https://www.sciencedirect.com/science/article/pii/S0167739X16308652}
\BIBentrySTDinterwordspacing

\bibitem{clevr}
J.~Johnson, B.~Hariharan, L.~van~der Maaten, L.~Fei-Fei, C.~L. Zitnick, and
  R.~Girshick, ``Clevr: A diagnostic dataset for compositional language and
  elementary visual reasoning,'' in \emph{2017 IEEE Conference on Computer
  Vision and Pattern Recognition (CVPR)}, 2017, pp. 1988--1997.

\bibitem{deLara20}
\BIBentryALTinterwordspacing
S.~P{\'{e}}rez{-}Soler, E.~Guerra, and J.~de~Lara, ``Model-driven chatbot
  development,'' in \emph{Conceptual Modeling - 39th International Conference,
  {ER} 2020, Vienna, Austria, November 3-6, 2020, Proceedings}, ser. Lecture
  Notes in Computer Science, G.~Dobbie, U.~Frank, G.~Kappel, S.~W. Liddle, and
  H.~C. Mayr, Eds., vol. 12400.\hskip 1em plus 0.5em minus 0.4em\relax
  Springer, 2020, pp. 207--222. [Online]. Available:
  \url{https://doi.org/10.1007/978-3-030-62522-1\_15}
\BIBentrySTDinterwordspacing

\end{thebibliography}
